# OQ Carinae – A New Southern Z Cam Type Dwarf Nova

**Rod Stubbings**
*Tetoora Road Observatory, 2643 Warragul-Korumburra Road, Tetoora Road 3821, Victoria, Australia; stubbo@sympac.com.au*

**Mike Simonsen**
*AAVSO, 49 Bay State Road, Cambridge, MA 02138; mikesimonsen@aavso.org*

ABSTRACT

Long term optical monitoring of the dwarf nova OQ Car has been conducted to study the previously unknown behaviour of this star system. The observations have shown OQ Car to have frequent dwarf nova outbursts and revealed the first recorded standstill of this star system. Based on this, we conclude that OQ Car is a new member of the Z Cam type dwarf novae.

INTRODUCTION

Z Camelopardalis (Z Cam) stars are known for random standstills midway between their outburst and quiescence state. The defining features show a fairly large degree of constant brightness, around 1 to 1.5 magnitudes below the outburst state. The standstills can last for a few days, months or years. The orbital periods of Z Cams are shown to be all over 3 hours (Simonsen et al. 2014).

Z Cam stars appear to be a relatively rare group as there are currently only 21 known confirmed systems (Simonsen, Bohlsen, Hambsch and Stubbings, 2014).

HISTORY

Spectroscopic observations established OQ Car to be a cataclysmic variable (CV). A dwarf nova classification was suggested due to the variation seen the UBVRI data (Cieslinski, Steiner and Jablonski. 1998). There was no period found up to 6 hours. This could be due to the low inclination of the system or the orbital period may exceed 6 hours. It was noted that the mean average magnitude from those observations was 15.8V (Woudt, Warner and Spark. 2005).

OQ Car is classified as a UG star in the International Variable Star Index (VSX) (Watson, Henden and Price. 2006) with a range of 14.5 – 17p. No previous long term monitoring of this object has been undertaken.

OBSERVATIONS

Close monitoring of the unstudied dwarf nova OQ Car commenced in July 2000 to establish the characteristics of this object. The observations have shown OQ Car to be a very frequently outbursting dwarf nova. Light curve analysis has shown an average outburst cycle of 14.9 days, which suggests little time is spent at minimum.

The average cycle time was determined using a custom designed tool in *VSTAR* (Benn 2012). Maxima were hand selected individually, and the average time between selections were automatically calculated. The averages of the densest data sets were then calculated to derive the mean time between selections.

The brightest maxima observed reach 13.6, which exceeds the VXS range of 14.5 – 17p. The minima were not observed directly in this study, as the visual observations only extend to a depth of 15.8 (Fig. 1).

In January 2014, following a normal outburst, OQ Car failed to return to its faint state and entered into a standstill (Fig. 2). The standstill, with a mean magnitude of 14.7 has lasted over 38 days so far.

CONCLUSION

The first long term optical monitoring of OQ Car over 14 years has revealed a very active dwarf nova with an outburst cycle of 14.9 days and an amplitude of 13.6v – 17p. The orbital period remains unknown. The January 2014 standstill is a unique feature that has not been previously observed. Approximately 1 magnitude fainter than maximum, the current standstill has well exceeded the outburst cycle of 14.9 days, lasting over 38 days so far. We conclude that this confirms OQ Car to be a Z Cam type dwarf novae.

ACKNOWLEDGEMENTS

This research has made use of the International Variable Star Index (VSX) database, operated at AAVSO, Cambridge, Massachusetts, USA. The authors wish to thank the anonymous referee whose comments and suggestions made this a better paper.

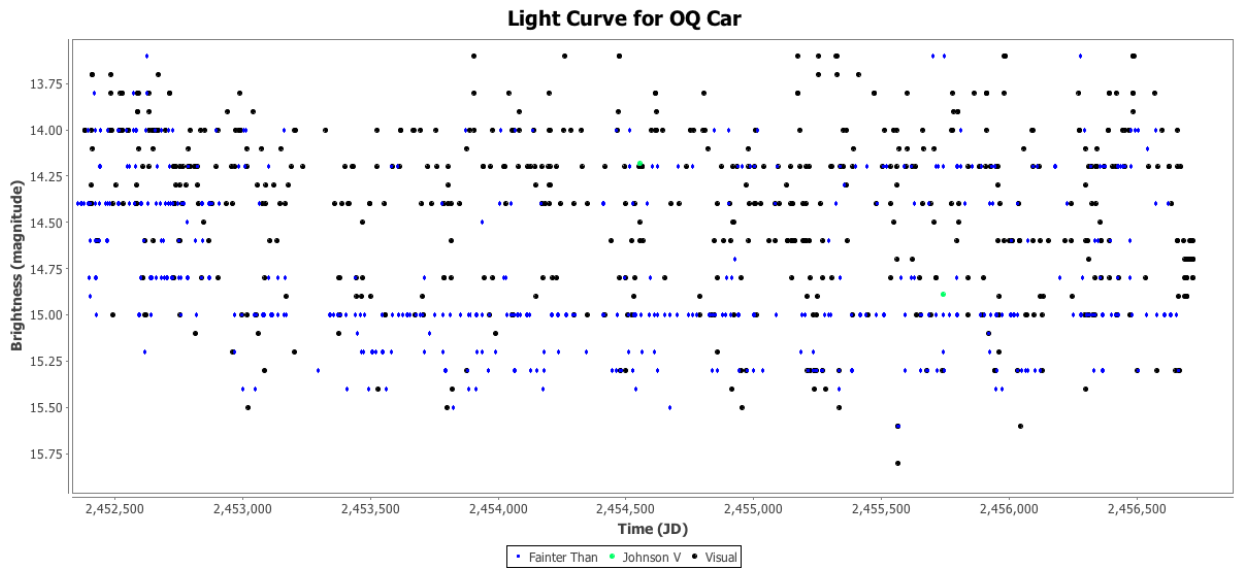

Figure 1. AAVSO historical data from JD 2452383 – 2456717 (April 18, 2002- March 1, 2014) showing the normal, active dwarf novae behavior of OQ Carinae. Black dots are visual data, green are Johnson V and blue are upper limits.

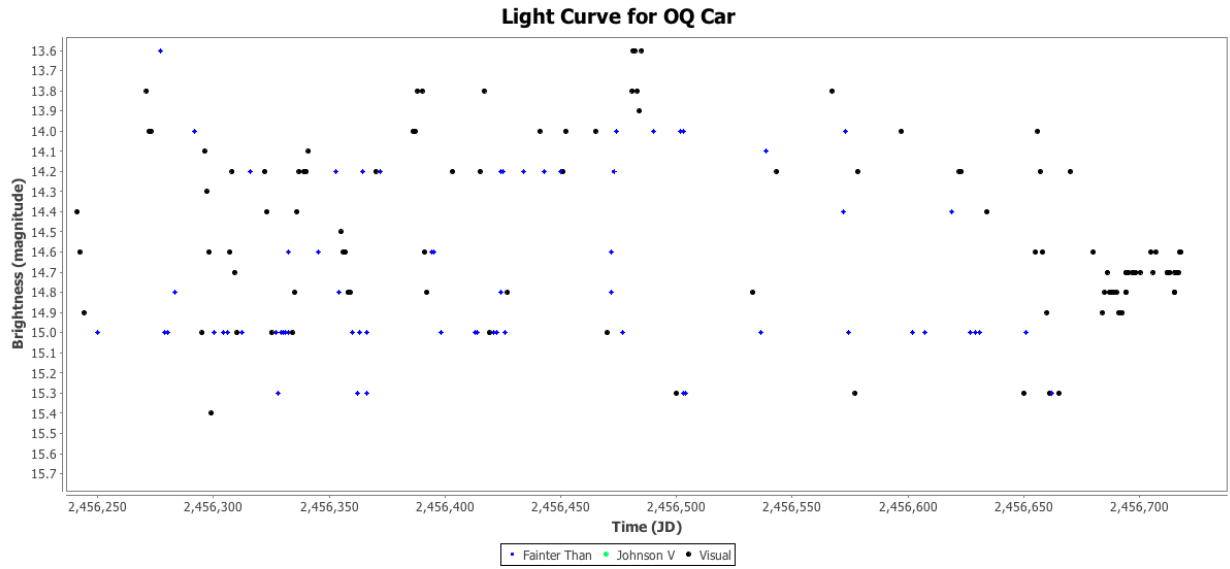

Figure 2. AAVSO visual data, showing the first known standstill of OQ Carinae (far right), beginning JD 2456679 through JD 2456717 (January 22, 2014 – March 1, 2014). Black dots are visual data, blue are upper limits.